\documentclass[journal,comsoc]{IEEEtran}
\usepackage{amsmath,amsfonts}
\usepackage{algorithmic}
\usepackage{algorithm}
\usepackage{array}
\usepackage[caption=false,font=normalsize,labelfont=sf,textfont=sf]{subfig}
\usepackage{textcomp}
\usepackage{stfloats}
\usepackage{enumerate}
\usepackage{url}
\usepackage{verbatim}
\usepackage{color}
\usepackage{graphicx}
\usepackage{cite}
\usepackage{balance}
\usepackage[justification=centering]{caption}
\begin{document}

\title{Cloud-Edge-Terminal Collaborative AIGC for Autonomous Driving}

\author{
    Jianan Zhang,
    Zhiwei Wei, 
    Boxun Liu,
    Xiayi Wang,
    Yong Yu,
    Rongqing Zhang
    \thanks{This work has been submitted to the IEEE Wireless Communications Magazine for publication. Copyright may be transferred without notice, after which this version may no longer be accessible.}
    \thanks{Jianan Zhang, Boxun Liu, Xiayi Wang, and Yong Yu are with the State Key Laboratory of Advanced Optical Communication Systems and Networks, School of Electronics, Peking University, Beijing 100871, China.}
    \thanks{Zhiwei Wei is with the Shanghai Research Institute for Intelligent Autonomous Systems, Tongji University, Shanghai 201210, China.}
    \thanks{Rongqing Zhang (corresponding author) is with the School of Software Engineering, Tongji University, Shanghai 200092, China, and also with the Shanghai Research Institute for Intelligent Autonomous Systems, Tongji University, Shanghai 201210, China.}
}

\maketitle

\begin{abstract}
    In dynamic autonomous driving environment, Artificial Intelligence-Generated Content (AIGC) technology can supplement vehicle perception and decision making by leveraging models’ generative and predictive capabilities, and has the potential to enhance motion planning, trajectory prediction and traffic simulation. This article proposes a cloud-edge-terminal collaborative architecture to support AIGC for autonomous driving. By delving into the unique properties of AIGC services, this article initiates the attempts to construct mutually supportive AIGC and network systems for autonomous driving, including communication, storage and computation resource allocation schemes to support AIGC services, and leveraging AIGC to assist system design and resource management.
\end{abstract}


\section{Introduction}
Autonomous driving has developed rapidly with the goal of improving traffic safety, efficiency, and convenience through perception, decision making, and control of vehicles. {Autonomous driving also involves interactions with human through communicating driving intentions and vehicle responses based on individual needs.} The technologies span multiple fields, such as computer vision, machine learning, sensor fusion and control theory, and is highly complex and challenging. {As autonomous driving systems progress from controlled environments and straightforward tasks to more complex and unpredictable urban landscapes, current technologies may falter to generalize and touch their performance limitations.}

Artificial Intelligence-Generated Content (AIGC) technology has the potential to empower autonomous driving. AIGC refers to the use of artificial intelligence technology to automatically or collaboratively generate various types of contents such as texts, images, audios, and videos based on user needs and goals. The core of AIGC technology is to use deep neural network models to learn the potential distribution of data and generate new data that conform to the distribution according to given conditions or goals. {This capability enables AIGC technology to generalize from learned data distributions to new scenarios.}
Language Foundation Models \cite{brown2020language} can be used to understand and generate natural languages, such as dialogue and summary, to improve the communication efficiency and quality between drivers and vehicles. Visual Foundation Models \cite{rombach2022high} can be used to detect and recognize objects, scenes, and emotions in images, improving perception of the surrounding environment. Multi-modal Foundation Models \cite{cherti2023reproducible} can be used to fuse different types of data, such as text, voice and  images, to improve driver’s entertainment experience and personalized needs. Moreover, AIGC can be applied to autonomous driving end-to-end. For example, DriveGPT4 \cite{xu2023drivegpt4}, a large-scale multi-modal language model for autonomous driving, uses videos collected by the vehicle and historical vehicle control decisions to output the next control decision, while providing natural language explanations for its decision to improve the interpretability.

Personalized data generation capabilities of AIGC enhance vehicle and driver copilot experience. First, AIGC can tailor the driving experience to the preferences and needs of each individual driver, such as adjusting the speed, route, and ambiance. Second, AIGC can provide personalized feedback and guidance to the driver, such as suggesting optimal driving habits, alerting potential hazards, and offering emergency assistance. Third, AIGC can enhance the communication and interaction between the driver and the vehicle, such as using natural language processing, voice recognition, and facial expression analysis. These features make autonomous driving more enjoyable, safe, and efficient for each driver.

However, stringent latency and availability requirements of autonomous driving tasks bring challenges of applying AIGC. A vehicle alone hardly has sufficient communication, storage, and computation resources to support large model storage and inference under a limited budget constraint, while model stored at the cloud requires communication between cloud and vehicle and has higher communication latency. To address this challenge, this article proposes a cloud-edge-terminal collaborative architecture and operation schemes to support different AIGC models and meet the quality-of-service requirements of autonomous driving tasks. 

The main contributions of this article are three-fold. First, we survey potential applications of AIGC to autonomous driving tasks. Second, we propose a cloud-edge-terminal collaborative AIGC architecture and service workflow to address the resource and latency challenges of applying AIGC to autonomous driving. Third, we initiate attempts to develop communication, storage, and computation resource allocation mechanisms to support the operation of the cloud-edge-terminal collaborative AIGC system, and leverage AIGC to enhance network design.

\section{AIGC for autonomous driving}
In this section, we first survey autonomous driving tasks that benefit from the generative and predictive capabilities of AIGC. We then discuss the challenges of applying AIGC to autonomous driving, which demands for location-dependent and personalized content under strict latency constraints. 

\subsection{AIGC Applications to Autonomous Driving}
\begin{figure*}
    \centering
    \includegraphics[width=6.2 in]{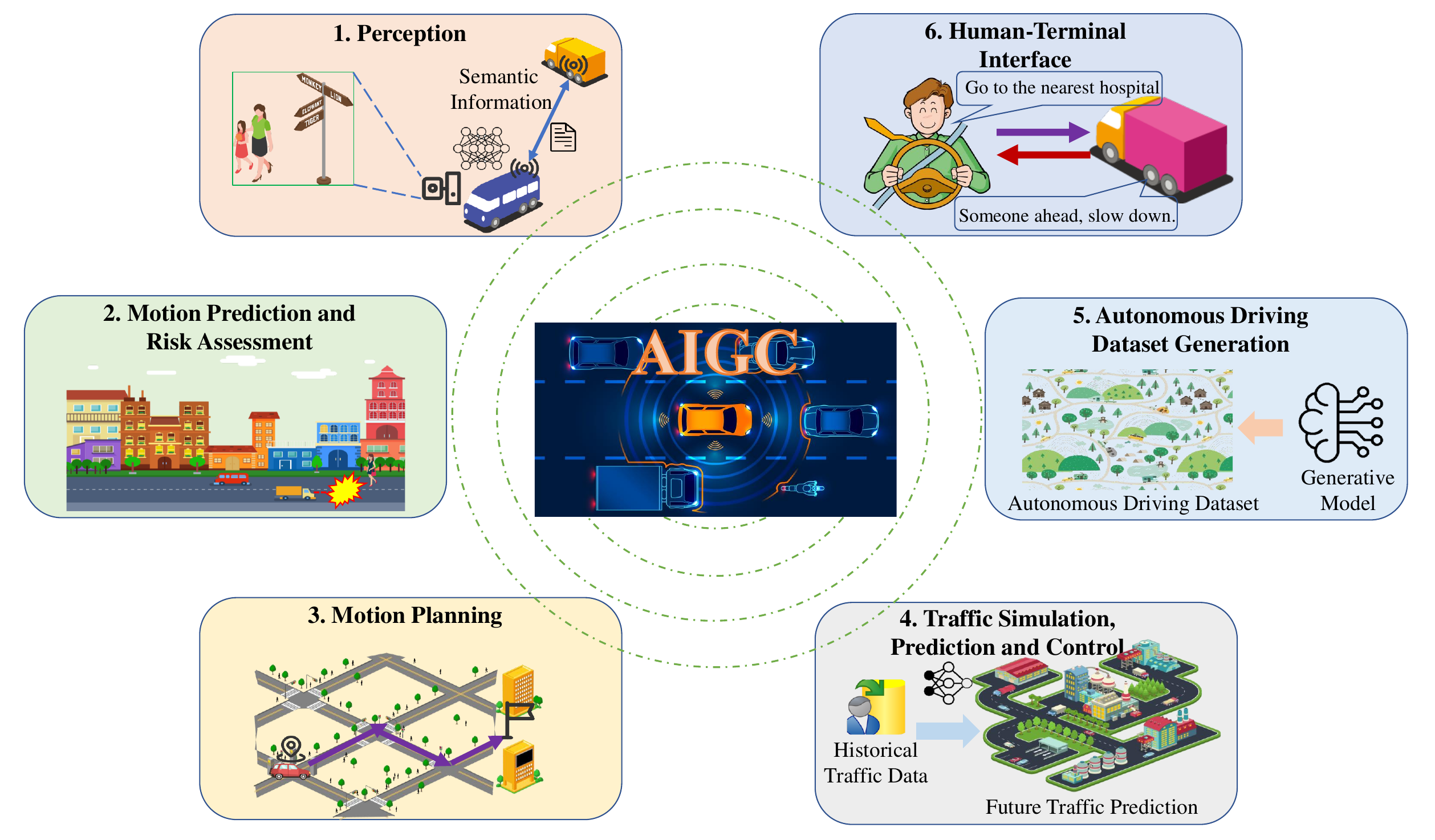}

    \caption{Applications of AIGC to autonomous driving.}
    \label{fig:applications}
\end{figure*}

AIGC can be applied to several aspects of autonomous driving, illustrated by Fig.~\ref{fig:applications}.

\textbf{Perception}: Perception is the process of acquiring, interpreting, and using sensory information to enable an autonomous vehicle to navigate safely and efficiently in complex and dynamic environments. 

Generative models can be used to supplement the characterization of unknown environments that are not present in sensory perceptions. For instance, generative models can estimate the likelihood of vehicles or pedestrians appearing behind an obstacle based on the perception near the obstacle. This is possible because generative models can recover a picture using only a small fraction of pixels \cite{rombach2022high}. Generative models can also infer terrain properties outside the perception range, which improves perception under cluttered environments where sensors are obstructed. Additionally, generative models can perform targeted object detection based on language prompts. For example, given a traffic scene image and a language prompt “detect traffic light colors,” the generative model can identify the traffic lights and provide corresponding text or voice descriptions to assist color-defective drivers.

Generative models can better extract semantics and reduce communication loads, while transmitting raw perceptual information requires a large amount of communication resources. For example, converting an image into a short text description and then transmitting the text description to other vehicles can reduce bandwidth and latency in cooperative sensing. 

\textbf{Motion prediction and risk assessment}: Motion prediction aims to estimate the trajectories of surrounding objects, such as vehicles, pedestrians and cyclists, based on their current and past positions, velocities, orientations, and other attributes. Risk assessment aims to predict collision and localize risky regions, given the predicted motions of the ego-vehicle and other moving objects. 

Generative AI models have the ability to learn the potential distribution of data, including the trajectories of vehicles and pedestrians under various scenarios. Moreover, the prediction task can be transformed into a language modeling task \cite{seff2023motionlm}, which can be used to predict the behaviors of vehicles and pedestrians. For example, given previous traffic scenes and a question “What will the last car in the middle lane do when the yellow light comes on ahead?” the language model can analyze the scene and provide a possible answer “The last car will accelerate through the intersection.” 

Based on motion prediction, generative models can perform risk assessments. For example, given an alleyway scene and traffic, the generative model will predict that “Two cars at the intersection ahead may collide.” 

\textbf{Motion planning}: Motion planning is the process of finding a feasible and safe trajectory for a vehicle to follow in a dynamic environment, including obstacle avoidance, lane keeping and path finding. {For example, the Tesla FSD V12 end-to-end autonomous driving model has been successfully applied in practical driving systems, achieving driving performance close to that of humans.}

Motion planning can also be transformed into a language modeling task \cite{mao2023gpt}. For example, given a traffic scene and a goal “from point A to point B,” the language model can generate a reasonable path and provide corresponding text descriptions \cite{zhang2023generative}.

In addition, generative language models can provide explanations for the decisions made. For example, given a traffic scene and a question “What should the autonomous vehicle do?” the model can generate an optimal behavior based on the scene and provide corresponding text explanations “The autonomous vehicle should slow down and stop on the side of the road because there is an ambulance passing ahead.”

\textbf{Traffic simulation, prediction and control}: Besides microscopic content generations such as vehicle’s local perception, motion prediction and planning, AIGC can be applied to macroscopic traffic control such as simulating, predicting and controlling traffic flows in complex urban environments \cite{xu2023generative}. 

By combining digital twins, generative models can simulate real driving environments in virtual space and evaluate traffic control strategies. Suppose that a control center has the knowledge of real-time traffic. Generative models at the control center can output traffic light control decisions that alleviate traffic congestion through simulations. The control decisions can then be implemented at traffic lights through communicating with Road Side Units (RSU). In addition, given a traffic scene in a city or region and a parameter “increase traffic flow by 10$\%$,” generative models can generate a new traffic scene though simulation, predict possible congestion situations, and provide corresponding text descriptions. 

\textbf{Autonomous driving dataset generation}: Autonomous driving requires high-quality data for training and testing. However, collecting and labeling real-world data is costly, time-consuming, and may not cover all possible scenarios. Therefore, synthetic data generation is a promising alternative that can provide large-scale, diverse, and high-fidelity data for autonomous driving applications.  

Generative models use existing data to create new traffic data that can provide more corner cases and various conditions for training autonomous driving algorithms. For example, given a traffic scene and a parameter “generate perception under rainy or snowy weather,” generative models can create new perception data by adding rain or snow effects to the original scene.

\textbf{Human-vehicle interface}: A human-vehicle interface is essential for ensuring the safety, comfort, and trust of the drivers and passengers in autonomous vehicles, since there are situations where human input or supervision may be required, such as in emergencies, complex scenarios, or legal regulations. The interface should be designed to provide clear and timely information, intuitive and easy-to-use controls, and adaptive and personalized features.

{Generative models enhance the way vehicles understand and respond to human input, including voices and gestures. Vehicle can anticipate the driver's needs and offer assistance proactively, making the driving experience more immersive and personalized. For instance, if the vehicle senses that the driver is tired or predicts that the driver will be tired based on the driving history, it can recommend a rest area or coffee shop ahead.}

Generative models can also explain the behavior of the vehicle, making it easier for humans to understand and supervise. For example, given a vehicle’s behavior “stop on the side of the road”, generative models can generate a text explanation based on the behavior “stop on the side of the road because there is an ambulance passing ahead” and provide feedback to the user through voice or screen.

\subsection{Challenges of Applying AIGC to Autonomous Driving}
\textbf{High computational complexity and latency}: To achieve safe autonomous driving, AIGC applications need to analyze and make decisions on the vehicle’s status, environment, road conditions, and other information in milliseconds. This poses a challenge to the computation and storage resources of vehicles because AIGC models usually have a large number of parameters and complex structures. If AIGC models are deployed in the cloud, network transmission delay, bandwidth limitation, and vehicular mobility issues may lead to a decreased quality of service. Edge collaboration is a feasible solution, which can use resources at RSU near the vehicle to provide low-latency, high-reliability, and high-performance AIGC services.

\textbf{Adaptive to different regions and traffic conditions}: Some autonomous driving tasks depend on geographical location and terrain environmental characteristics. Different countries or regions may have different road rules and traffic regulations, such as driving directions, speed limit signs and traffic lights. These rules will affect the behavior selection and content generation of autonomous driving vehicles in different regions. For example, in the United States, autonomous driving vehicles need to drive on the right, while in the United Kingdom, they need to drive on the left. On highways, autonomous driving vehicles need to adjust their speed according to speed limit signs, while on city roads, they need to stop or start according to traffic lights. In mountain roads, obstacles are complex and have various shapes, while the vehicle’s field of view is limited, requiring more cautious decision-making. Therefore, AIGC services need to be able to identify localized rules and properties in different regions and generate appropriate content based on these rules.

Additionally, vehicles may encounter varied road conditions and traffic flows, which necessitate adapting navigation strategies accordingly. For example, in congested road sections, autonomous driving vehicles may need to change lanes or decelerate more frequently, while in open road sections, they can accelerate or maintain stability. In these cases, AIGC services need to be able to perceive the current system environment and generate content that is highly adaptive, flexible, and efficient based on this environment.

\textbf{Personalized content for ego-vehicles}: Different vehicle owners may have different driving habits and preferences. For example, some people like to drive smoothly, while others like to drive fast; some people like to maintain a certain distance, while others like to follow closely; some people like to change lanes in advance, while others like to change lanes temporarily. These preferences will affect the decision-making mode of autonomous driving vehicles, such as when, where, and how to accelerate, decelerate, change lanes, and overtake. Therefore, AIGC services need to be able to learn and adapt to the personalized preferences of vehicle owners and generate content that meets their preferences.

Vehicle owners may also have different style preferences for generated content. For example, some people like concise and clear content, while others like detailed and rich content; some people like formal and rigorous content, while others like humorous and relaxed content. These preferences will affect the content expression of autonomous driving vehicles, including voice prompts, image displays, and text displays. Therefore, AIGC services need to be able to recognize and adapt to the personalized styles of vehicle owners and generate content that meets their preferences.

Providing personalized vehicle control and content style to the driver poses a significant challenge for generative models, as they have to generate tailored responses based on a limited amount of interaction history, while the driving history may span a long period of time. Therefore, generative models need to learn how to efficaciously capture the preferences, habits, and goals of the driver from a few conversational turns or metadata of the vehicle and the driver, and generate relevant and coherent content that adapts to the driving context.

\begin{figure*}
    \centering
    \includegraphics[width=6.2 in]{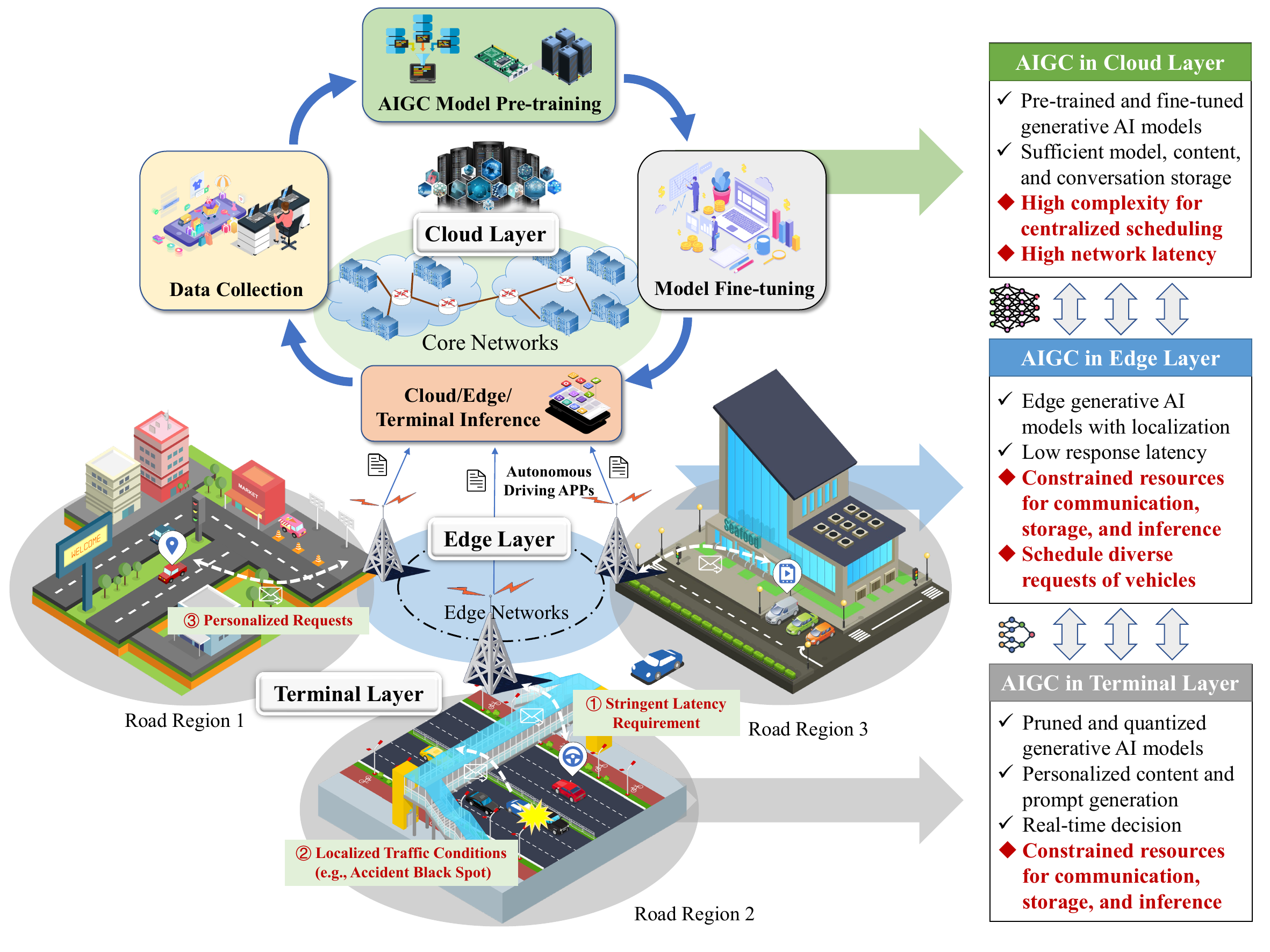}

    \caption{The cloud-edge-terminal collaborative AIGC architecture.}
    \label{fig:cloud_edge_terminal}
\end{figure*}
\section{Cloud-Edge-Terminal Collaborative AIGC Architecture}
We propose a cloud-edge-terminal collaborative AIGC architecture to support low-latency, location-dependent, and personalized autonomous driving task requests. Based on the architecture, we further discuss the mutual assistance between AIGC services and network resource management. We initiate attempts of leveraging AIGC to improve network communication, storage and computation resource management, and proposing resource allocation schemes to support AIGC for autonomous driving.

\subsection{Architecture Design}
The cloud-edge-terminal collaborative AIGC architecture is shown in Fig. \ref{fig:cloud_edge_terminal}.

\textbf{Cloud}: AIGC Service Providers (ASP) train large generative models using a large amount of data, and deploy pre-trained and fine-tuned models with strong reasoning capabilities by utilizing sufficient computing and storage resources in the data center. These models handle complex AIGC tasks such as high-quality traffic simulations, predictions, and traffic control evaluations. ASP can also customize large models according to the characteristics and needs of different regions and compress them into smaller models for deployment at the edge.

\textbf{Edge}: At the edge, RSU obtains smaller fine-tuned generative models from the cloud and schedules resources to provide localized and timely responses for vehicles. Most AIGC services for autonomous driving can be completed at the edge, including perception, motion prediction, and risk assessment. Only when the smaller models at the edge are insufficient to complete the specified task, the task is offloaded to the cloud. For example, RSU can collect local traffic information and send it to the cloud for centralized traffic light control using larger models.

\textbf{Terminal}: Vehicles host pruned and quantized generative models using limited computation and storage resources. The models generate personalized content using lightweight computation and therefore have limited capabilities \cite{li2020gan}. To support more demanding generative tasks, a vehicle chooses the appropriate ASP at the edge or cloud, sends the request to the ASP, and the generated content is then communicated back to the vehicle. The generated content can be further processed at the terminal to meet personalized needs. To enhance the personalization of content generation, a long interaction history can be compressed semantically \cite{jiang2023llmlingua} and stored in the user profile. 

\textbf{Service workflow} in cloud-edge-terminal collaborative AIGC architecture for autonomous driving includes the following steps.

1. AIGC service request generation: {The vehicle or driver at the terminal generates a request}, which may include a personalized prompt. {The system decides whether to execute locally or offload to edge (or cloud)}, by considering task complexity, privacy, and latency requirements.

2. ASP selection and offloading: {Given computation and storage resource constraints of different ASPs at the edge (or cloud), and communication constraints between the vehicle and the edge (or cloud), vehicles select the appropriate ASPs for their task requests.} The objective of ASP selection is to generate high-quality responses within the resource and latency constraints. The generated content can then be transmitted from the remote ASP to the vehicle.

3. Post-processing of generated content: Some content returned by remote ASP may be in intermediate formats to reduce the communication load (e.g., features or text description of an image) and {need to be further processed by generative models at the terminal to be consumed by the driver}. Moreover, post-processing can further personalize the content since generative models at the vehicle maintain a more comprehensive set of driver preferences.

While the above service workflow focuses on model inference and content generation, the large number of requests and traffic information can further enhance model training and fine-tuning. By analyzing additional traffic data, ASP can update the generative models at the cloud and the edge. This process is latency insensitive and requires large amounts of resources, in contrast with real-time service requests.

\subsection{System Operations}
AIGC relies on communication, storage, and computation resources that collaborate throughout the whole workflow. Compared with traditional workloads, AIGC services have unique properties that can be incorporated in the design of network resource allocation strategies. First, the versatile content generation capability can adapt not only to users, but also to the available network resources. Generative models are able to output contents of various qualities (e.g., pictures of various resolutions). Satisfying resource demands of generation tasks with available network resources through joint optimization of task adaptations and resource allocations has the potential to increase user satisfaction and utility. {Second, different from traditional content distribution networks where users request the same content stored at edge servers, AIGC services generate content tailored to the terminal users, and the generated content could even change in response to the same question of the same user as interaction evolves. Therefore, AIGC services require more tightly coupled computation and storage resources in the collaborative framework.} On one hand, the stored model requires computation resources to generate content, and caching alone is insufficient to provide personalized content. On the other hand, content generation depends on the interaction history, which occupies additional storage resources, and the interaction is unique to the user, in contrast to the traditional content distribution network where different users send the same request. Therefore, additional storage resources are needed to support users’ stateful service requests, whereas traditional requests are often stateless. In the remainder of this section, we discuss task adaptation and resource allocation strategies to support AIGC services, and leverage AIGC to assist resource management.

\begin{figure*}
    \centering
    \includegraphics[width=5 in]{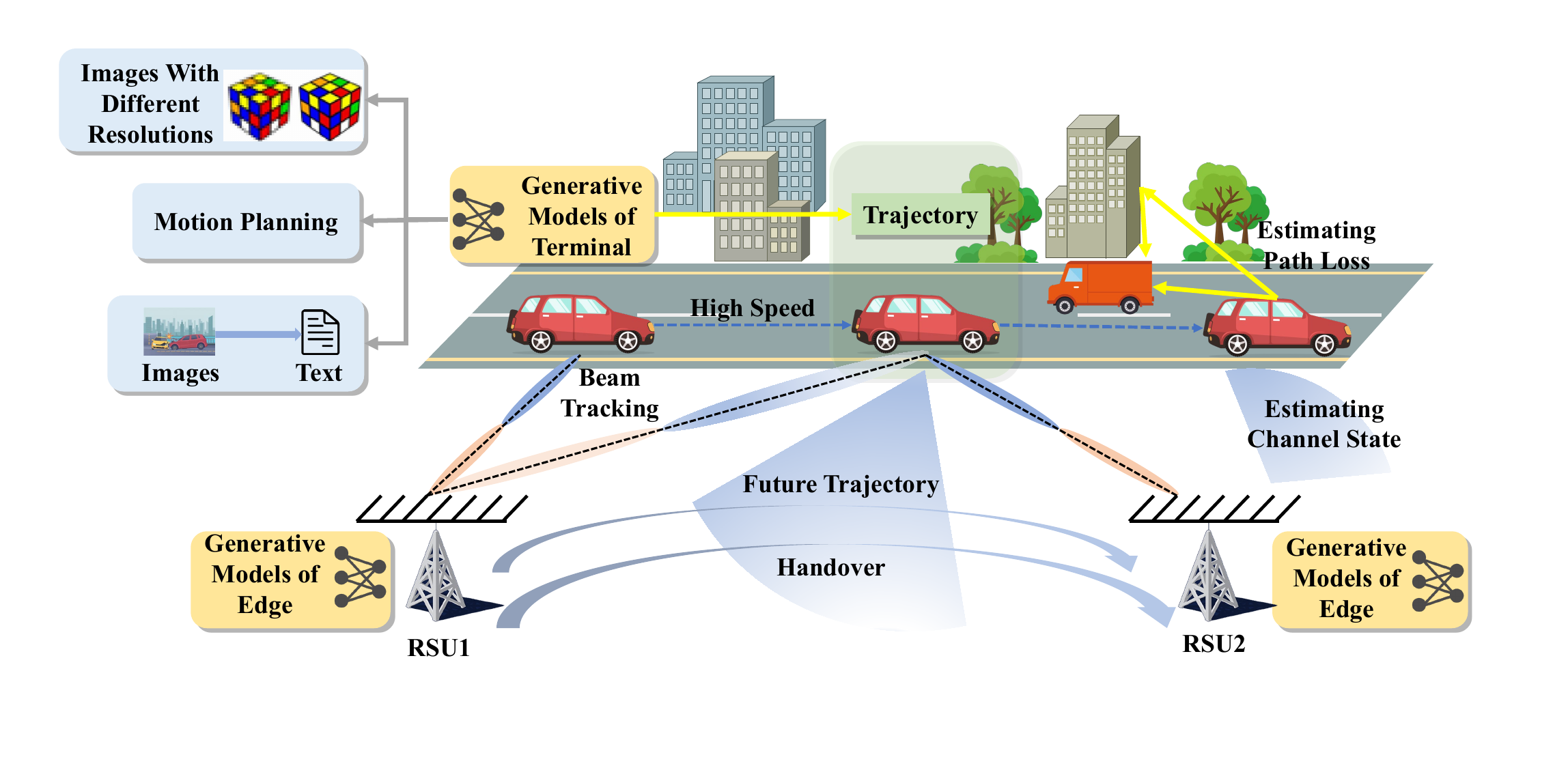}
    \caption{{AIGC for elastic task generation, communication resource prediction and allocation.}}
    \label{fig:communication}
\end{figure*}

\textbf{Communication:} One of the challenges of autonomous driving is to ensure reliable and efficient communication between vehicles and infrastructures when vehicles are moving at high speeds. Link transmission capacity varies rapidly over time, especially for higher frequency band wireless communications such as 6G, which can affect the quality and timeliness of the data exchanged. {Addressing this challenge involves two issues: determining what to transmit and ensuring efficacy under rapidly changing network conditions. Elastic task generation and resource allocation match task demands and network resources in dynamic vehicular networks, and can both be enhanced by applying AIGC technologies.}

{On one hand, generative models can be used to create content with different volumes that match the current link transmission capacity.} For example, a generative model can produce a low-resolution image when the link is weak, and a high-resolution image when the link is strong. A generative model can also convert images to texts, which further reduces data volumes for communications. This way, the generative models can adapt to the changing network conditions and optimize data transmission for autonomous driving. 

{On the other hand, AIGC has the potential to improve wireless communications by addressing mobility challenges and proactively allocating resources.} Generative models for motion planning can predict the future trajectories of vehicles based on their previous states and the environment. As shown in Fig. \ref{fig:communication}, the predicted vehicle locations can improve beam tracking between vehicles and access points in RSU or base station. Higher frequency band wireless transmission is prone to blockage and scattering. Environment perception improves beam forming by identifying the locations and types of blockers and scatterers \cite{SoM}. Moreover, predicting vehicle trajectories facilitates handover, in ultra-dense small cells with shorter transmission range and line-of-sight transmission for 6G, by selecting the optimal access point to communicate with the vehicles. The network can proactively reserve bandwidth in the most probable adjacent cell instead of all adjacent cells to reduce bandwidth waste before the handover occurs, to improve resource utilization and handover accuracy. 

{Generative models can also assist communication resource allocation by predicting dynamic communication demand and available bandwidth resources.} AIGC for macroscopic traffic simulation estimates the future traffic flow intensity, which reflects the amount of communication resources required by the vehicles. AIGC for microscopic motion planning and perception can assist in estimating channel state information by using sensing information and reducing pilot overhead. These predictions can then be used to improve the routing and scheduling algorithms that manage the communication networks and enhance their performance. 
{By applying AIGC to task generation, resource prediction and allocation, the cloud-edge-terminal framework can optimize the utility of communication operations and satisfy various task demands.}

\begin{figure*}
    \centering
    \includegraphics[width=6.2 in]{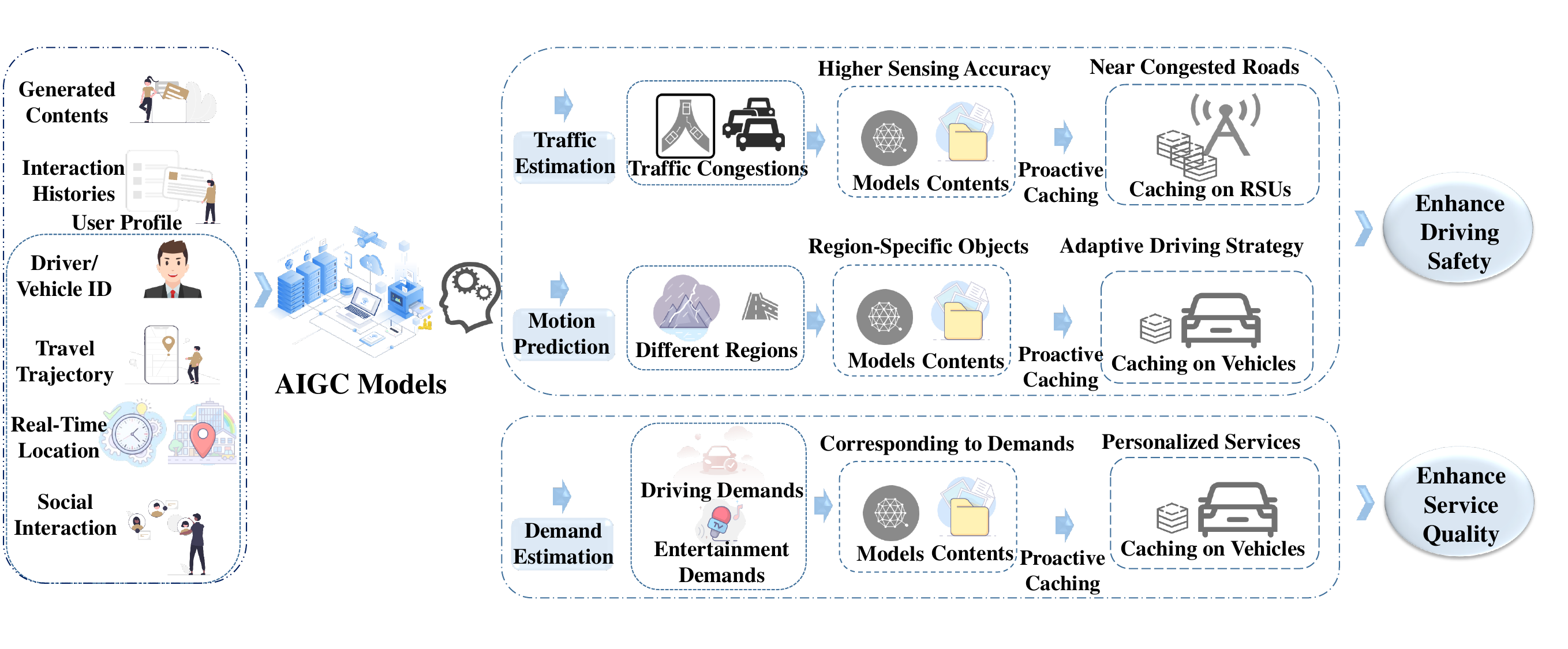}
    \caption{Proactive caching based on AIGC estimation and prediction for autonomous driving.}
    \label{fig:Storage}
\end{figure*}

\textbf{Storage:} Contents, models, user profiles, and interaction histories can all be stored since they are key ingredients for AIGC services. Caching generated content avoids repeated model inference computation, and can serve multiple vehicles that have the same request. However, in the case of diverse service requests and preferences in autonomous driving, to obtain personalized content and increase the cache hit rate, caching models become necessary \cite{xu2023joint}. A model can generate different content for different users according to user profiles and their interaction histories. {A user profile includes driver identity, vehicle type information, travel trajectory, real-time location and speed, social interaction with other vehicles, etc., reflecting users’ driving behavior and entertainment preferences.} Interaction history is recorded while using AIGC service, such as input prompts and responses, reflecting users’ preferences and feedback on system performance. Both user profiles and interaction histories can be compressed to save storage space. Moreover, popular generative language models cannot generalize to longer texts than the training sequence lengths. Compressed contents or features extracted from the interaction history can instead be used when interacting with generative models.  

Since vehicles and RSU have limited storage resources, they cache most relevant models and content by estimating future demand, shown in Fig. \ref{fig:Storage}. Proactive caching based on AIGC traffic estimation and motion prediction enhances driving safety while reduces service latency. With traffic congestion prediction, relevant models and content with higher sensing accuracy can be cached in RSU near congested road sections, to meet vehicles’ navigation demand on higher-risk congested roads while reduces service latency and network load. As vehicles drive across different regions, region-specific generative models can be proactively cached in vehicles. For example, before entering mountainous terrains, higher precision perception models can be proactively cached. Such models facilitate identifying different types of obstacles in mountainous terrains and inferring potential obstacles outside vehicle’s perception range, which assists vehicles in complex environment to take a more conservative driving strategy. In mountainous terrains, where RSUs are limited, proactive caching in vehicles becomes necessary.

In addition, proactive caching based on user demand estimation enhances service quality. Using user profiles and interaction histories as input, generative models can predict future driving and entertainment demands. The corresponding models and contents can be cached in advance. For instance, generative models with recreation features can be proactively cached in a vehicle before a road trip, which can be inferred based on the interaction history.

\begin{figure*}
    \centering
    \includegraphics[width=6.2 in]{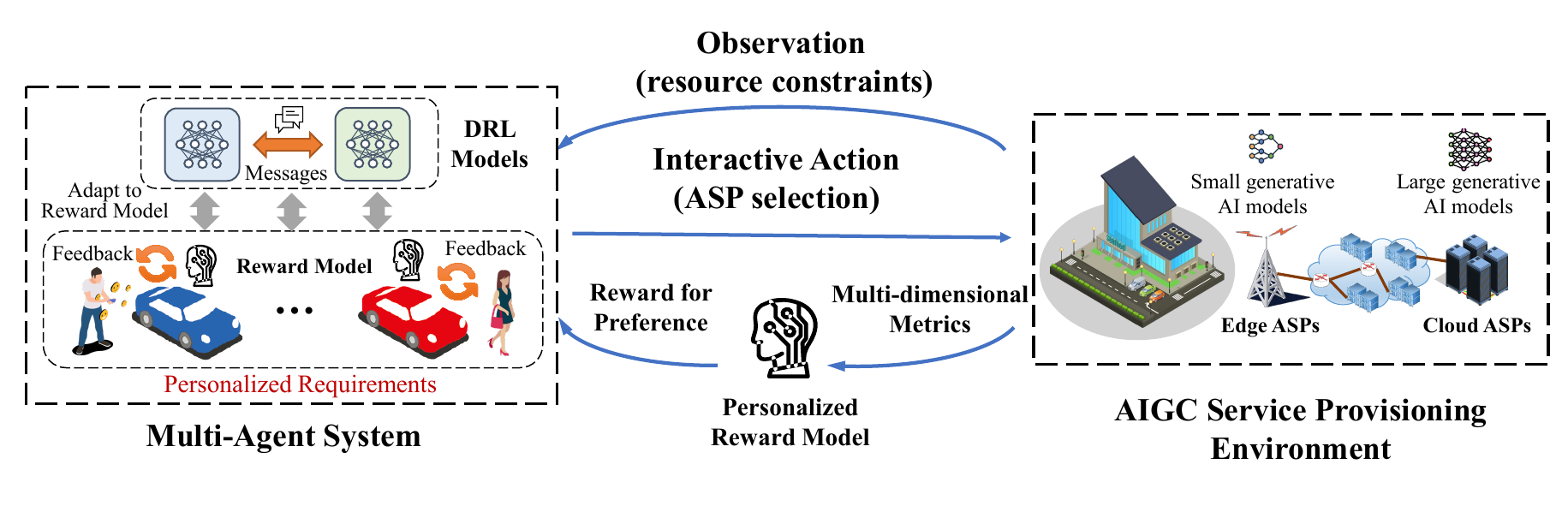}
    \caption{The interactive MARL framework for ASP selection based on drivers' personalized reward models.}
    \label{fig:compuatation}
\end{figure*}

\textbf{Computation:} Computation is the core of the cloud-edge-terminal collaborative AIGC architecture to support real-time personalized AIGC services. The challenges of allocating computation resources for autonomous vehicles are two-fold. On one hand, it is difficult to quantify the subjective and personalized preferences of users towards the generated content. Consider a tourist in an autonomous vehicle who prefers scenic routes over the fastest ones. The system must understand the subjective concept of “scenic value”, which might include views and landmarks. This preference varies greatly from one individual to another and depends on personal taste and current mood. On the other hand, the multi-user resource allocation problem for multiple ASPs, i.e., the ASP selection, is framed as a resource-constrained task assignment problem \cite{xu2023unleashing}, which is NP-hard \cite{tindell1992allocating}. The complexity of this problem is further exacerbated when considering the autonomy of vehicles.

To align the autonomous systems with human preferences, the first step involves training a cross-modal semantically capable model to serve as the Reward Model (RM). This model is crucial for understanding and interpreting the multi-modal data streams pertinent to AIGC tasks, and is fine-tuned using human feedback collected from long-term interactions with generative models as well as user profiles. This feedback mechanism allows the model to capture a wide range of human preferences, from aesthetic considerations in content generation to pragmatic concerns like energy efficiency and latency. The RM processes input parameters such as the quality of generated content, latency, and energy consumption, and produces a scalar reward as an output. This reward represents a quantified estimation of user satisfaction, encapsulating the complex and subjective nature of human preferences in a format that is computationally manageable for optimization.

Vehicles generate multiple AIGC service requests at different times, which will be processed by ASPs. The ASP selection problem is complex due to the autonomy of vehicles, diverse user preferences, resource constraints, dynamic vehicular environments, and interdependence of selection decisions. Given these features, the Partially Observable Stochastic Game (POSG) framework is anticipated to capture the intricacies of this decision-making process. POSG accounts for the partial observability inherent in each vehicle’s decision-making process, where a vehicle may not have complete information about the network state or the service choices being made by other vehicles. Moreover, POSG accommodates the dynamic nature of the environment and the strategic interplay among multiple vehicles, each striving to optimize its own service experience in the context of limited resources at the edge. To navigate the challenges posed by the POSG framework, we propose an interactive Multi-Agent Reinforcement Learning (MARL)\cite{iqbal2019actor} approach for ASP selection, shown in Fig. \ref{fig:compuatation}. This approach allows vehicles to engage in pre-decision interactions with nearby vehicles through communication modules. Such interactions enable vehicles to share insights and coordinate their decisions, thereby enhancing the overall decision-making process in the context of ASP selection. The actions, rewards, and human feedback are stored for continuous training and refinement of the RM, thus ensuring that the system remains adaptable and responsive to the evolving user preferences.

In summary, by integrating the RM concept into the AIGC task management framework for autonomous vehicles, we provide a robust MARL mechanism to align automated decision-making processes with the nuanced and dynamic nature of human preferences. This approach not only enhances the relevance and effectiveness of AIGC tasks but also ensures that autonomous systems remain user-centric and adaptable in real-world scenarios.

\section{Conclusion}
Mobility in complex autonomous driving environments poses challenges to vehicle perception and decision-making. Generative models can augment perception and predict future vehicle motions by leveraging the generative capability based on distributions learned from previous data. This article surveys the potential applications of AIGC to autonomous driving and proposes a cloud-edge-terminal collaborative architecture to support AIGC. The unique properties of generative models bring challenges to communication, storage, and computation resource allocations, while models’ predictive capability can assist network design and resource management. This article delves into the challenges and research opportunities, and proposes initial attempts to construct mutually supportive AIGC and network systems for autonomous driving. 

\section{Acknowledgments}
This work was supported in part by the National Natural Science Foundation of China under Grants 62341101, 62301011, and 62271351.


\end{document}